\documentclass[a4paper,twocolumn,floatfix,showpacs,showkeys,amsmath,amssymb,nobibnotes,altaffilletter]{revtex4-1}
\usepackage{graphicx}
\usepackage{pslatex}
\usepackage{xspace}
\usepackage{subfigure}

\usepackage{tabularx}
\usepackage{multirow}
\usepackage{booktabs}

\usepackage[pdfauthor={Jens Lorrmann},%
pdftitle={Charge Carrier Extraction by Linearly Increasing Voltage: Analytic framework and ambipolar transients},%
pagebackref=false,
pdfkeywords={CELIV, analytical solution, ricatti, charge extraction by linearly increasing voltage},
pdftex]{hyperref}
\hypersetup{
    colorlinks,%
    citecolor=black,%
    filecolor=black,%
    linkcolor=black,%
    urlcolor=black
}

\newcommand{\bfig}{\begin{figure}}
\newcommand{\efig}{\end{figure}}
\newcommand{\bit}{\begin{itemize}}
\newcommand{\eit}{\end{itemize}}

\newcommand{\pht}{poly(3-hexyl thiophene-2,5-diyl)\xspace}
\newcommand{\pcbm}{[6,6]-phenyl-C$_{61}$ butyric acid methyl ester\xspace}

\newcommand{\etal}{\emph{et\,al.}\xspace}
\newcommand{\PCLV}{photo-CELIV\xspace}
\newcommand{\dfr}[3][ ]{\frac{ {\rm d}^{#1}#2}{ {\rm d}#3^{#1}}}
\newcommand{\dott}{\textperiodcentered}
\newenvironment{figref}[1]{Fig.~\ref{#1}}{}
\newenvironment{eqnref}[1]{Eqn.~(\ref{#1})}{}

\graphicspath{{figs/}{bilder/pdf/}}

\begin{document}

\preprint{}

\title{Charge Carrier Extraction by Linearly Increasing Voltage:\\Analytic framework and ambipolar transients}

\author{J.~Lorrmann$^1$}\email{jens.lorrmann@physik.uni-wuerzburg.de}
\author{B.~H.~Badada$^2$}
\author{O.~Ingan{\"a}s$^2$}
\author{V.~Dyakonov$^{1,3}$}
\author{C.~Deibel$^1$}\email{deibel@physik.uni-wuerzburg.de}
\affiliation{$^1$Experimental Physics VI, Julius-Maximilians-University of W{\"u}rzburg, 97074 W{\"u}rzburg, Germany}
\affiliation{$^2$Biomolecular and Organic Electronics, IFM, Center of Organic Electronics, Link{\"o}ping University, S-5813 Link{\"o}ping, Sweden}
\affiliation{$^3$Bavarian Center for Applied Energy Research e.V. (ZAE Bayern), 97074 W{\"u}rzburg, Germany}

\date{\today}

\begin{abstract}
  Up to now the basic theoretical description of charge extraction by linearly increasing voltage (CELIV) is solved for a low conductivity approximation only. Here we present the full analytical solution, thus generalize the theoretical framework for this method. We compare the analytical solution and the approximated theory, showing that especially for typical organic solar cell materials the latter approach has a very limited validity. Photo-CELIV measurements on poly(3-hexyl thiophene-2,5-diyl):[6,6]-phenyl-C$_{61}$ butyric acid methyl ester based solar cells were then evaluated by fitting the current transients to the analytical solution. We found that the fit results are in a very good agreement with the experimental observations, if ambipolar transport is taken into account, the origin of which we will discuss. Furthermore we present parametric equations for the mobility and the charge carrier density, which can be applied over the entire experimental range of parameters.
\end{abstract}

\maketitle

\section{Introduction}
  Bulk heterojunction solar cells use a phase separated blend of an electron accepting and electron donating material --- e.g., \pht~(P3HT) and \pcbm~PCBM --- as active layer~\cite{deibel2010review}. In this nano-scale blended film the photo-generated charges are separated at the donor-acceptor interface and collected at the electrodes under short circuit conditions. State-of-the-art polymer-based solar cells provide high yields for collected charges with respect to the incident photons~\cite{park2009} and have reached a power conversion efficiency (P.C.E.) up to $7.9\%$ under AM1.5 ($\mathrm{100 mW/cm^2}$) illumination~\cite{liang2010,green2010review}.

  The knowledge about the physical processes, such as recombination and charge carrier transport, and their impact on the charge collection in organic solar cells is crucially important for an optimization of the P.C.E and has therefore been intensively debated in literature~\cite{deibel2009c,coropceanu2007review}. The recombination and the charge transport are governed by the charge carrier lifetime $\tau$ and the charge carrier mobility $\mu$, respectively. Various techniques have been used to study charge carrier dynamics in these systems, e.g. time-of-flight photo-conductivity~\cite{baumann2008} for the transport or transient absorption~\cite{de2007,clarke2009a,shuttle2008a} for the recombination. Another approach to measure charge carrier mobility and recombination was introduced about ten years ago by G. J\v{u}ska~\etal~\cite{juska2000}, called CELIV, charge carrier extraction by linearly increasing voltage. Due to its ability to measure these two parameters simultaneously this technique has attracted much interest in the organic semiconductor research. However, the theory for calculating the current response due to the linearly increasing voltage has only been presented for the simplified cases of low and high conductivity regimes, respectively~\cite{juska2000}. As we will see later, to our knowledge neither the low conductivity approximation $\tau_\sigma = \epsilon \epsilon_0/ en\mu \gg t_{tr}$ ($\tau_\sigma$ is the relaxation time, $t_{tr}$ is the time at which all free charge carriers are extracted) nor the high conductivity approximation $\tau_\sigma \ll t_{tr}$ is valid for state-of-the-art organic solar cell materials, such as P3HT and PCBM.

  In this paper we present the full analytical framework for the CELIV method which allows us to computationally evaluate the mobility $\mu$ and the  charge carrier density $n$ of experimental measurements. In general, a closed analytical expression for the relation $\mu(t_{max})$ cannot be derived, as for the high or the low conductivity approximation. Therefore we propose a parametric equation for the mobility evaluation from CELIV experiments which is valid over the entire experimental range of conductivities. An equivalent expression is derived for the charge carrier density $n(t_{max})$. The mobility equation is tested against parametric mobility equations known from the literature~\cite{juska2000,juska2004,bange2010,deibel2009c}. Furthermore we evaluate experimental photo-current transients by fitting the data directly to the CELIV framework. \\

\section{CELIV theory}
\subsection{Method summary}
\begin{figure}[t]
    \centering
    \includegraphics[width=.45\textwidth]{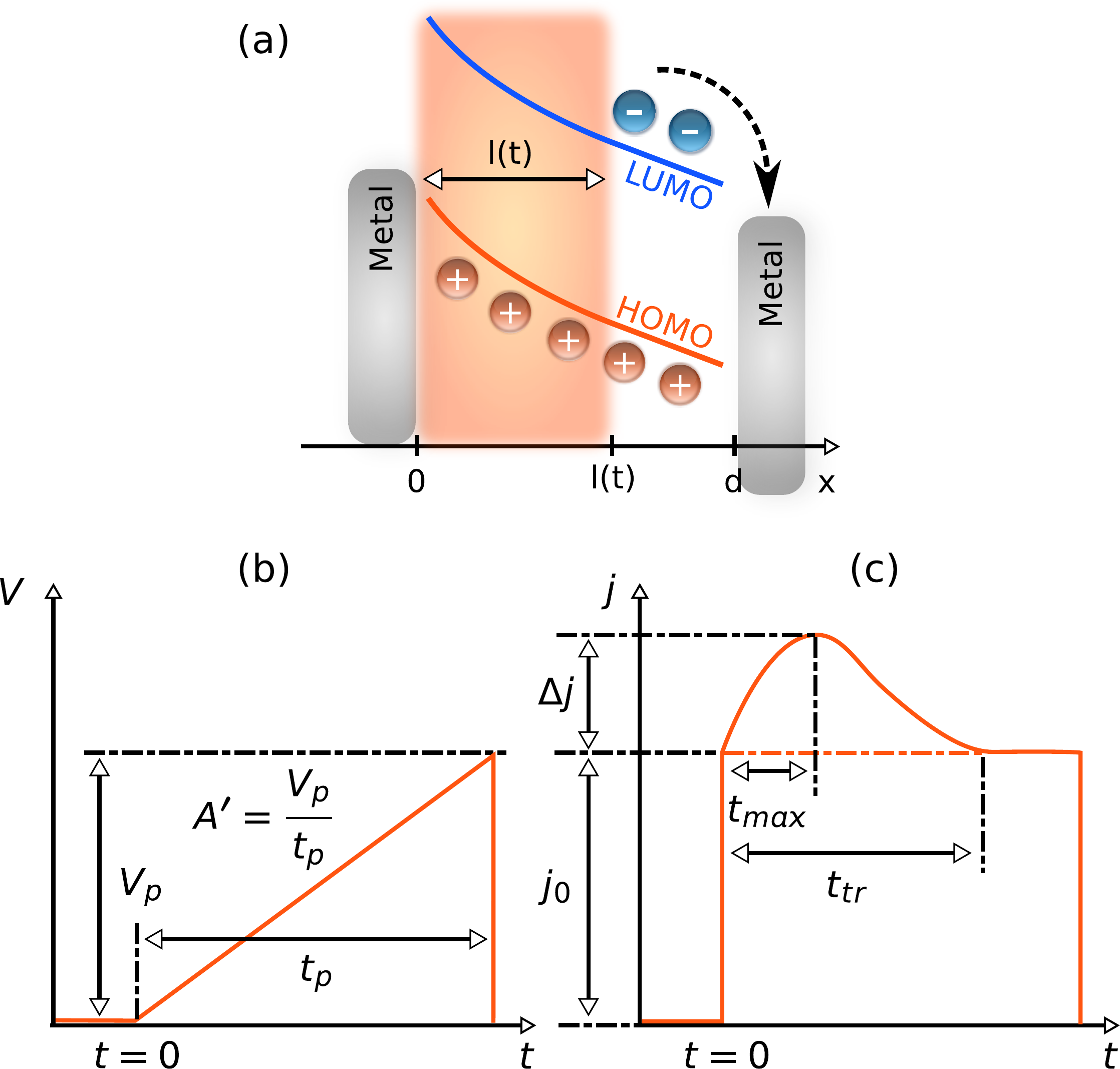}\hfill
    \caption{(colour online) Schematic illustration of the CELIV method. (a) Process of charge extraction and the band diagram in the active material sandwiched between two electrodes at a certain time. The region $0 \leq x \leq l(t)$ where all electrons have already been extracted is depicted by the shaded area. (b) Scheme of the voltage input and (c) the current output. The voltage pulse with slope $A'$ is applied in reverse bias. This yields a characteristic current density response with a capacitive offset $j_0$ and current density due to the drift of the free charge carriers with a maximum current of $\Delta j$.}
    \label{fig:CELIV_Scheme}
\end{figure}

A schematic illustration of the CELIV technique is given in \figref{fig:CELIV_Scheme}. In this method a linearly increasing voltage $V(t)=A't$, where $A'$ is the slope of the applied voltage pulse, is used to extract equilibrium charge carriers with density $n$ and mobility $\mu$ from a film with a certain dielectric permittivity $\epsilon$ and thickness $d$. The whole device is represented as a capacitor with the film between two electrodes at $x=0$ (blocking contact) and $x=d$. We here summarize the important parts of the CELIV theory~\cite{juska2000} for comprehension and later discussions.

It is assumed that one charge carrier is much more mobile than the other (unipolar transport), that the electrode dimensions are much larger than the device thickness and the charge carrier density of free charge carriers is $n$. This yields the following charge carrier distribution $\rho(x,t)$ at time $t$
\begin{eqnarray}
    \rho(x,t) = \begin{cases}
        en  &, 0 \leq x \leq l(t)\\
        0   &, x > l(t)
        \end{cases}
    \label{eq:1}\mathrm{~.}
\end{eqnarray}
This expression is related to the time dependent extraction depth $l(t)$, where $0 \leq l(t) \leq d$. At a certain time $t$  in the region $0 \leq x \leq l(t)$ all electrons have been extracted and the region is charged positively.

Applying the increasing voltage $V(t)$ with the condition $RC \ll t$ the total current density $j(t)$ in the external circuit due to the redistribution of the charge carriers (electric field) is
\begin{subequations}
\label{eq:3}
\begin{eqnarray}
    j(t) &=& j_0 + \rho(t)\left( \frac{\mu A'}{d} t - \frac{en\mu}{2\epsilon\epsilon_0 d}l(t)^2\right)\mathrm{~~,with}\label{eq:3c}\\
    j_0&=&\frac{\epsilon\epsilon_0A'}{d}\label{eq:3a}\\
           \rho(t)&=&
           \begin{cases}
                en\left(1-\frac{l(t)}{d}\right)&,0 \leqq l(t) \leqq d\\
                0  &, d < l(t) \\
           \end{cases}\label{eq:3b}\mathrm{~.}
\end{eqnarray}
\end{subequations}
$R$ and $C$ are the device resistance and the device capacitance, respectively. $j_0$ is the differentiating initial step of the RC-circuit, $\rho(t)$ is the density of free charges in the device and becomes zero when $l(t)=d$. The last term in brackets in~\eqnref{eq:3c} describes the drift of the free charge carriers due to the external field and due to the electric field caused by the charge distribution in the sample. In the simple case the drift due to the latter electric field is neglected. For the calculation of the current transient $j(t)$ the extraction depth $l(t)$ is the crucial parameter. $l(t)$ can be expressed as a Ricatti-type first order differential equation
\begin{eqnarray}
    \dfr{l(t)}{t} = -\frac{en\mu}{2\epsilon\epsilon_0d}l(t)^2 + \frac{\mu A'}{d}t
    \label{eq:4}\mathrm{~.}
\end{eqnarray}
This equation has the initial conditions
\begin{eqnarray}
  l(0)&=0\\
  \dfr{l(t)}{t}|_{t=0}&=0
\end{eqnarray}
 and has up to now either been solved only numerically~\cite{bange2010} or analytically for a low ($\tau_\sigma\gg t_{tr}$) and a high ($\tau_\sigma\ll t_{tr}$) conductivity approximation~\cite{petravicius1976,juska2000}, respectively. Note that in the low conductivity approach \eqnref{eq:4} simplifies to ${\rm d}l(t)/{\rm d}t=\mu A't/d$ and thus the extraction depth $l(t)$ becomes
\begin{eqnarray}
    l(t)=\frac{\mu A' t^2}{2d}
    \label{eq:5}\mathrm{~.}
\end{eqnarray}
From \eqnref{eq:5} the transit time $t_{tr}$ --- corresponding to $l(t)=d$ when all free charge carriers are extracted --- can be defined as
\begin{eqnarray}
    t_{tr}=d\sqrt{2/\mu A'}
    \label{eq:5a}\mathrm{~.}
\end{eqnarray}
In the next section we introduce the analytical solution for the Ricatti-type definition of the extraction depth $l(t)$ (\eqnref{eq:4}). By doing this, the CELIV theory is generalized and can be compared with the low conductivity approach. This way we can prove that the low conductivity approximation has a very limited validity for state-of-the-art disordered organic semiconductors used in organic electronic devices such as organic solar cells.


\subsection{Solving the Ricatti equation}
For solving \eqnref{eq:4} directly we need to use the substitution
\begin{eqnarray}
    \dfr{L(t)}{t}= \frac{en\mu}{2\epsilon\epsilon_0d} l(t)\cdot L(t)
    \label{eq:6}\mathrm{~.}
\end{eqnarray}
\eqnref{eq:6} is again a differential equation for the substituent $L(t)$. Inserting \eqnref{eq:6} in \eqnref{eq:4} transforms the latter equation to a Stokes type differential equation of second order
\begin{eqnarray}
 \dfr[2]{L(t)}{t} -\frac{en\mu}{2\epsilon\epsilon_0d}\frac{\mu A'}{d}t\cdot L(t) = 0
  \label{eq:8}\mathrm{~.}
\end{eqnarray}
This equation has two linearly independent solutions --- the Airy functions of first kind $Ai(x)$ and second kind $Bi(x)$~\cite{airy1838}. The linear combination of these two functions has to be normalized due to the boundary conditions. Finally we re-substitute \eqnref{eq:6} to the final analytic form of $l(t)$. For a more detailed description we refer to the Appendix and the referenced literature~\cite{abramowitz1954book,olver1974}. Thus l(t) is defined as
\begin{subequations}
\label{eq:9}
\begin{eqnarray}
    l(t)&=& \frac{\mu A' }{d\chi^2}t^2\frac{\sqrt{3} Ai'\left(\chi\right) + Bi'\left(\chi\right)}{\sqrt{3} Ai\left(\chi\right) + Bi\left(\chi\right)}\mathrm{,}
  \label{eq:9a}\\
  \chi&=&\left(\frac{en\mu}{2\epsilon\epsilon_0d}\frac{\mu A'}{d} t^3\right)^{1/3}\label{eq:9b}\mathrm{~.}
\end{eqnarray}
\end{subequations}
It is interesting to note that the first order of the series expansion of \eqnref{eq:9} is exactly the $l(t)$ of the low conductivity solution~(\eqnref{eq:5}) as used in Ref.~\cite{juska2000}.

\eqnref{eq:9} provides a complete analytical description of the CELIV framework. Unfortunately, the argument of the Airy functions in $l(t)$ involves the parameters mobility $\mu$ and charge carrier density $n$. Therefore it is not possible to algebraically extract an expression for these two parameters, as it is possible for the mobility using a low or high conductivity approximation~\cite{juska2000}. However, we show later in  Sec.~\ref{sec:parametric} that the analytical framework can be used to derive two parametric equations for $\mu$ and $n$. Furthermore, this analytical model is capable of determining the experimental parameters from \PCLV experiments by fitting the photo-current transient to \eqnref{eq:3}. Unfortunately the Airy function of second kind $Bi(x)$ and its derivative rises steeply even for small values of the argument $x>0$ and the fitting routine quickly gets computational cancellation problems. To solve this issue we first do a scaling of the involved parameters to render the whole framework dimensionless (see Sec.~\ref{sec:scaling}) and, secondly, use the confluent hyper-geometric function $_0F_1$~\cite{abramowitz1954book,andrews2000book} to represent the Airy functions (see App.~\ref{hypergeo}).

\subsection{Scaling to dimensionless parameters}
\label{sec:scaling}

To yield a dimensionless system, the extraction depth $l(t)$ is related to the sample thickness $d$ and the time $t$ is divided by the low conductivity case transit time $t_{tr}$, \eqnref{eq:5a}. This results in
\begin{eqnarray}
  \tilde{l}&=&d^{-1}l\label{eq:scale0}\\
  \tilde{t}&=&\frac{en\mu}{2\epsilon\epsilon_0}\sqrt{\tilde{A}'}t\label{eq:scale1}\mathrm{~.}
\end{eqnarray}
Hence the scaled device thickness is $\tilde{d}=1$. Furthermore, the scaling for $\tilde{j}$ and $\tilde{A}'$ is set to yield a Ricatti equation (\eqnref{eq:4}) and an extraction current density equation (\eqnref{eq:3}) which is parametric in the dimensionless voltage slope $\tilde{A}'$ only.
\begin{eqnarray}
  \tilde{A}'&=& \frac{2\epsilon^2\epsilon_0^2}{e^2n^2d^2\mu}A'\label{eq:scale2}\\
  \tilde{j}&=& \frac{2\epsilon\epsilon_0}{e^2n^2d\mu}j\label{eq:scale3}\\
  \dfr{\tilde{l}}{\tilde{t}}&=&-\frac{1} {\sqrt{\tilde{A}'}}\tilde{l}^2-2\tilde{t}\label{eq:scale4}\\
  \tilde{j}&=&\tilde{A}'+\left(1-\tilde{l}\right) \left(2\sqrt{\tilde{A}'}\tilde{t}-\tilde{l}^2\right)\label{eq:scale5}
\end{eqnarray}

\begin{figure}[t]
      \centering
      \includegraphics[width=.45\textwidth]{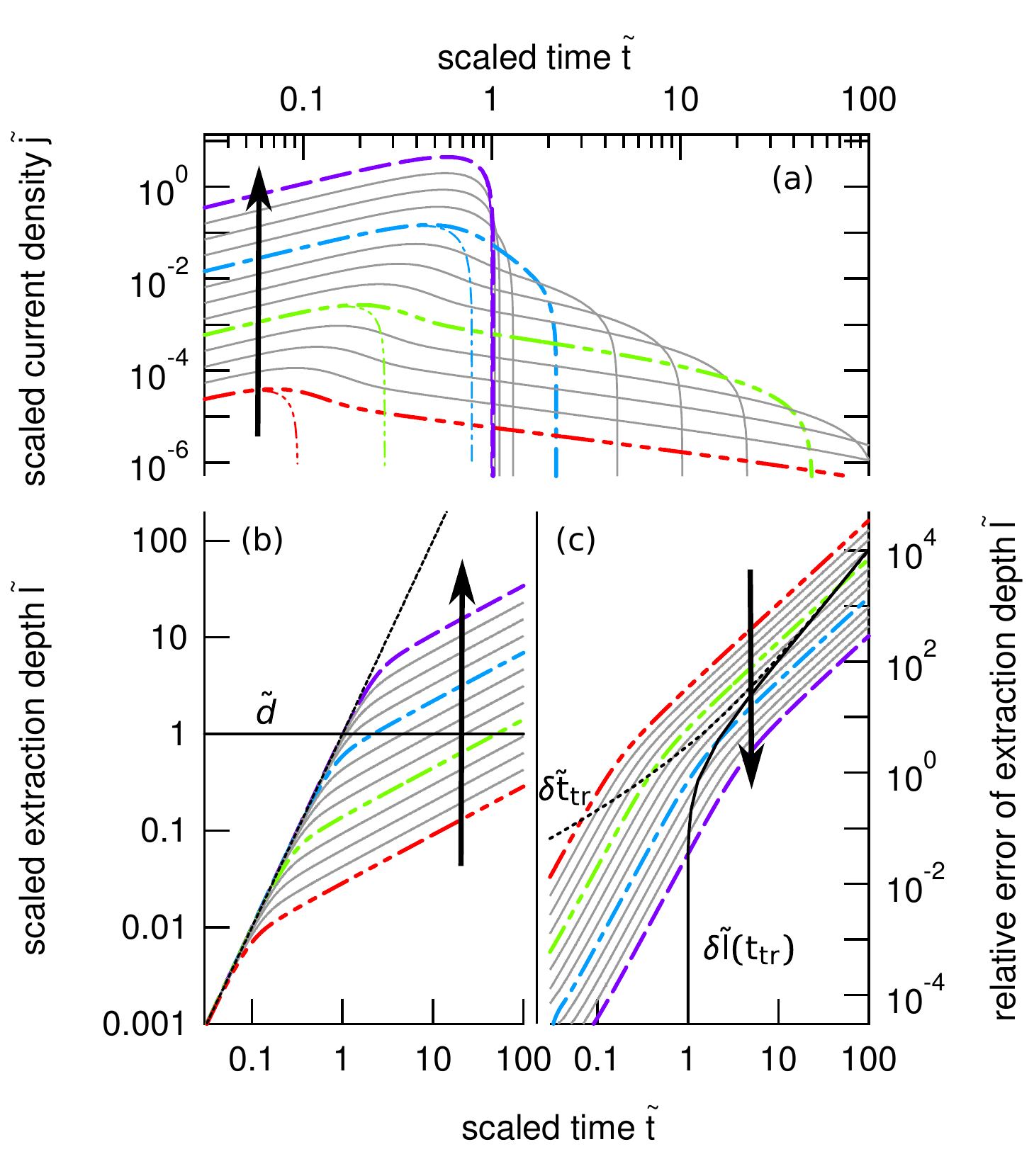}\hfill
      \caption{Results for the calculation of (a) the dimensionless extraction current $\tilde{j}$, (b) the dimensionless extraction depth $\tilde{l}$ and (c) its relative error $\delta l$ in case of the low conductivity case. The dimensionless voltage slope $A'$ rises between $-6.64 \leq \log(\tilde{A}') \leq 1.36$ for the different transients indicated by the black arrow in each of the sub-figures. More precise the bold dash-dotted lines corresponds to: (~-~-~) $\log(\tilde{A}')=1.36$, (~-~\dott~-~)  $\log(\tilde{A}')=-1.30$, (~-~\dott~\dott~-~)  $\log(\tilde{A}')=-3.94$, (~-~\dott~\dott~\dott~-~)  $\log(\tilde{A}')=-6.64$.\\ (a) The thin lines are the calculated transients for the low conductivity approximation and the bold ones are the results from the general scaled CELIV framework. (b) The black dotted line represents the low conductivity approximation and the bold ones \eqnref{eq:9}. As a guide to the eye the black solid horizontal line marks the dimensionless device thickness $\tilde{d}$. (c) The black solid line gives the relative extraction depth error at the time of extraction $\delta l(t_{tr})$ and the black dashed line displays the relative transit time error time $\delta t_{tr}$ on the x-axis.}
      \label{fig:fig1}
  \end{figure}

This scaling has two advantages --- first it enables the computation of CELIV extraction currents over a wide range of parameters and second it allows us to illustrate the difference between the low conductivity case and the exact case in a clear manner (see \figref{fig:fig1}).

Note that from \eqnref{eq:scale1} and \eqnref{eq:scale2} the following expression for the mobility $\mu$ and the charge carrier density $n$ are found
 \begin{eqnarray}
    \mu=&\frac{\displaystyle 2d^2}{ \displaystyle A' t_{max}^2}\tilde{t}_{max}^2\label{eq:mu_scale}\\
    n=&\frac{\displaystyle \epsilon_0\epsilon A' t_{max}}{\displaystyle e d^2}\frac{\displaystyle 1}{\displaystyle \tilde{t}_{max}\tilde{A}'^{0.5}}\label{eq:n_scale}\mathrm{~.}
\end{eqnarray}

These expressions depend on parameters which are not experimentally accessible, such as $\tilde{t}_{max}$ in \eqnref{eq:mu_scale} and $\tilde{t}_{max}\tilde{A}'^{0.5}$ in \eqnref{eq:n_scale} respectively. To overcome this issue we can use the scaled system and calculate the current density $\tilde{j}(\tilde{t})$ for a set of parameters. From the resulting transients we can relate some of its characteristics, namely $\tilde{t}_{max}$, $\Delta \tilde{j}$, $\tilde{j}_0$ and $\tilde{A}'$, to each other. All of these depend on $\tilde{A}'$ only because of the scaling, which yields a definition of $\tilde{j}(\tilde{t})$~(\eqnref{eq:scale5}) which is parametric in $\tilde{A}'$ only. These relations between the characteristics and $\tilde{A}'$ are displayed in \figref{fig:fig4}~(c) (see Ref.~\cite{juska2000, juska2000a} for comparison). In Sec.~\ref{sec:parametric} we show how to derive parametric approximations for $\tilde{t}_{max}$ and $\tilde{t}_{max}\tilde{A}'^{0.5}$ therefrom, which transform \eqnref{eq:mu_scale} and \eqnref{eq:n_scale} to depend just on experimentally accessible parameters. However, we compare the generalized solution with the low conductivity approach of the CELIV framework first.

\subsection{Comparing generalized analytical and low conductivity CELIV framework}

  In \figref{fig:fig1} we used the scaling to compare the low conductivity and generalized case regarding the extraction depth $\tilde{l}(\tilde{t})$ and the extraction current density $\tilde{j}(\tilde{t})$ on a logarithmic time scale parametric in the dimensionless voltage slope $\tilde{A'}$. The black arrow indicates the direction of increasing $\tilde{A'}$. \figref{fig:fig1}~(a) demonstrates scaled extraction current density transients $\tilde{j}(\tilde{t})$. The thin lines were calculated with the low conductivity approach and the bold without it.

  Due to the scaling of the time axis, in case that the transient drops at $\tilde{t}=1$ the low conductivity approximation and the analytical solution are almost equal. The other transients show a clear distinction, with a long extraction tail for the general analytical solution. This long tail turns up when the charge carrier drift due to the external field and due to the redistribution of the internal field  in \eqnref{eq:3c} are balanced ($A'/t \approx enl(t)^2/\left(2\epsilon\epsilon_0\right)$). In that case the shape of the dimensionless extraction depth $\tilde{l}(t)$ turns from a parabola $\tilde{l} \propto \tilde{t}^2$ to $\tilde{l} \propto \sqrt{\tilde{t}}$.

  This is demonstrated in \figref{fig:fig1}~(b). Here the thin dotted straight line is the extraction depth $\tilde{l}$ calculated from \eqnref{eq:5} (low conductivity), the buckled bold curves represent \eqnref{eq:scale0} (general solution). Hereby this straight line has a slope of $2$ for all values of $\tilde{A}'$. The general solution qualitatively follows this shape depending on $\tilde{A}'$ up to a certain time and then deviates from the low conductivity approximation. The quantitative deviation is plotted in \figref{fig:fig1}~(c) in terms of the relative error  of the low conductivity extraction depth $\delta\tilde{l}(\tilde{t})=\delta l(t)$. The solid black line in \figref{fig:fig1}~(b) indicates the scaled device thickness $\tilde{d}$. When $\tilde{l}(\tilde{t})=\tilde{d}$ all charges have been extracted from the device.

  The black solid line in \figref{fig:fig1}~(c) represents the relative error $\delta l(t_{tr})$ at time $\tilde{t}$ when $\tilde{l}(\tilde{t})=\tilde{d}$. From the time of intersection $\tilde{t}_{int}$ of this black solid line with any of the bold lines the relative transit time error can be calculated: $\delta t_{tr}=\tilde{t}_{int}-1$. The relative error in transit time $\delta t_{tr}$ is represented by the black dotted line which shows $\delta l(\delta t_{tr})$ --- $\delta l$ on the y-axis versus $\delta t_{tr}$ on the x-axis.

  From this two curves we see that especially for negative values of $\log(\tilde{A}')$ the relative error $\delta l(t_{tr})$ gets very large, as well as the error in transit time $\delta t_{tr}$. Furthermore, we can estimate the boundary of the low conductivity case validity and thus of the mobility equation derived from it (Eqn.~(11) in Ref.~\cite{juska2000}). Setting the limit for the required transit time error to $\delta t_{tr} \le 10\%$ we can derive the following condition for the dimensionless voltage slope $\tilde{A}'$ from the scaled framework
  \begin{eqnarray}
      \tilde{A}'>1\mathrm{.}
      \label{eq:11}
  \end{eqnarray}
  For instance, in \figref{fig:fig1}~(c) the second grey solid line from the bottom is $\tilde{A}'= 1.12$ and the time at intersection $\tilde{t}_{int}=1.12$. Hence the error is $\delta t_{tr}\approx 10\%$. With values for common experimental and material parameters ($A'=8\cdot10^4~\mathrm{V/s}$, $d=105~\mathrm{nm}$, $n=6\cdot10^{22}~\mathrm{m^{-3}}$ and $\epsilon=3.7$)~\cite{deibel2008,deibel2009c} $\tilde{A}'>1$ corresponds to a mobility $\mu < 1.69\cdot10^{-6}~\mathrm{cm^2/Vs}$.

\subsection{Deriving parametric equations}
\label{sec:parametric}
\begin{figure}[t]
    \centering
    \includegraphics[width=.5\textwidth]{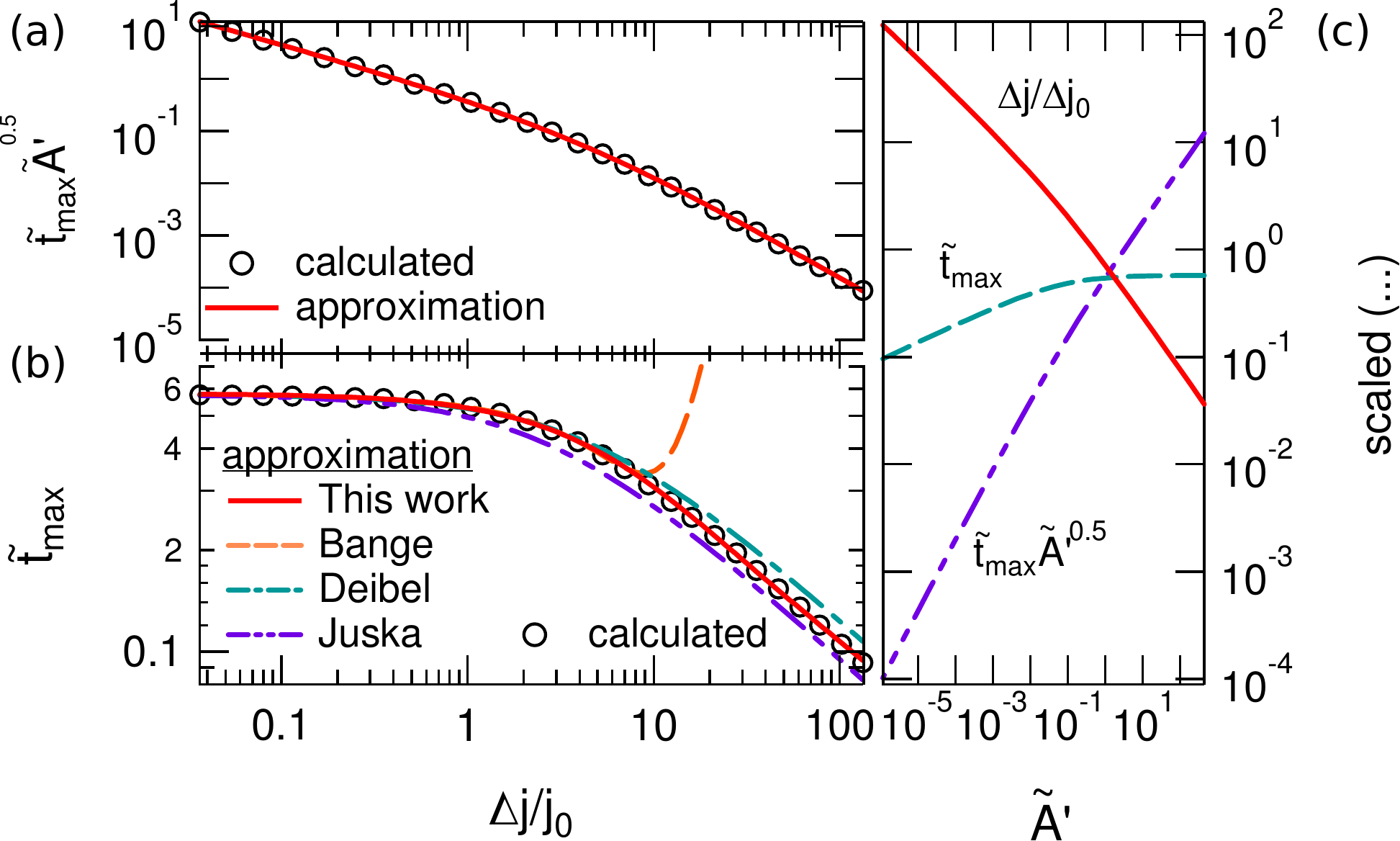}\hfill
    \caption{Calculated relation and parametric approximations for (a) $\tilde{t}_{max}\sqrt{\tilde{A}'}$ and (b) $\tilde{t}_{max}$ on $\Delta j/j_0$. (c) Overview over the relation between the involved scaled characteristic parameters $\Delta \tilde{j}/\tilde{j}_0$ (solid line), $\tilde{t}_{max}$ (dashed line), $\tilde{t}_{max}\tilde{A}'^{0.5}$ (dash dotted line) and the scaled voltage slope $\tilde{A}'$.}
    \label{fig:fig4}
\end{figure}

  The dimensionless parameters in \eqnref{eq:mu_scale} and in \eqnref{eq:n_scale} inhibit the determination of the mobility $\mu$ and the charge carrier density $n$ from experiments. However, the scaled dimensionless CELIV framework affords the opportunity to derive parametric equations for the charge carrier mobility $\mu$ and the charge carrier density $n$ from the characteristics of CELIV transients, in particular from $\tilde{t}_{max}$ and $\tilde{t}_{max}\tilde{A}'^{0.5}$  (see \eqnref{eq:mu_scale}, \eqnref{eq:n_scale}). Whose evolution with varying dimensionless voltage slope $\tilde{A}'$ is shown in~\figref{fig:fig4}~(c) together with $\Delta \tilde{j}/\tilde{j}_0$, which equals the ratio of the real unscaled parameters $\Delta j/j_0$. From~\figref{fig:fig4}~(c) $\tilde{t}_{max}$ and $\tilde{t}_{max}\tilde{A}'^{0.5}$ can be related to $\Delta j/j_0$. Anyhow, analytical definitions for these relations are not available and therefore we approximate them by means of parametric definitions. This finally yields expressions for the mobility $\mu$ and the charge carrier density $n$ including experimental parameters only. In the following we present the parametric results we determined. Furthermore, we compare the mobility equation~\eqnref{eq:mulorr} with parametrizations previously suggested in literature~\cite{juska2000,juska2000a,bange2010,deibel2009c}.

  In \figref{fig:fig4}~(a) the calculated relation between $\tilde{t}_{max}\tilde{A}'^{0.5}$ and $\Delta j/j_0$ (black circles) and the parametric equation we found (red line) are compared. The calculated curve has a slope of $-1$ for small values of $\Delta j/j_0 < 1$ and larger values yield a slope $\approx -2$. To describe this evolution we found a parametrization that describes this shape with a root mean square deviation $\sigma=0.8\%$,
  \begin{eqnarray}
      \tilde{t}_{max}\tilde{A}'^{0.5}=0.455\frac{j_0}{\Delta j}\left(1+0.238\frac{\Delta j}{j_0}\right)^{-1.055}
      \label{eq:scalen}\mathrm{~.}
  \end{eqnarray}
  From \figref{fig:fig4}~(a) it becomes clear that this approximation yields reasonable good fits. The relative error for all values is smaller than $2\%$. By means of \eqnref{eq:n_scale} and \eqnref{eq:scalen} the following equation can be used for the determination of the charge carrier density
  \begin{eqnarray}
      n=\frac{\epsilon_0\epsilon A' t_{max}}{0.455\cdot e d^2}\frac{\Delta j}{j_0}\left(1+0.238\frac{\Delta j}{j_0}\right)^{1.055}\mathrm{~.}
      \label{eq:nlorr}
  \end{eqnarray}

\begin{figure}[t]
  \centering
  \includegraphics[width=.5\textwidth]{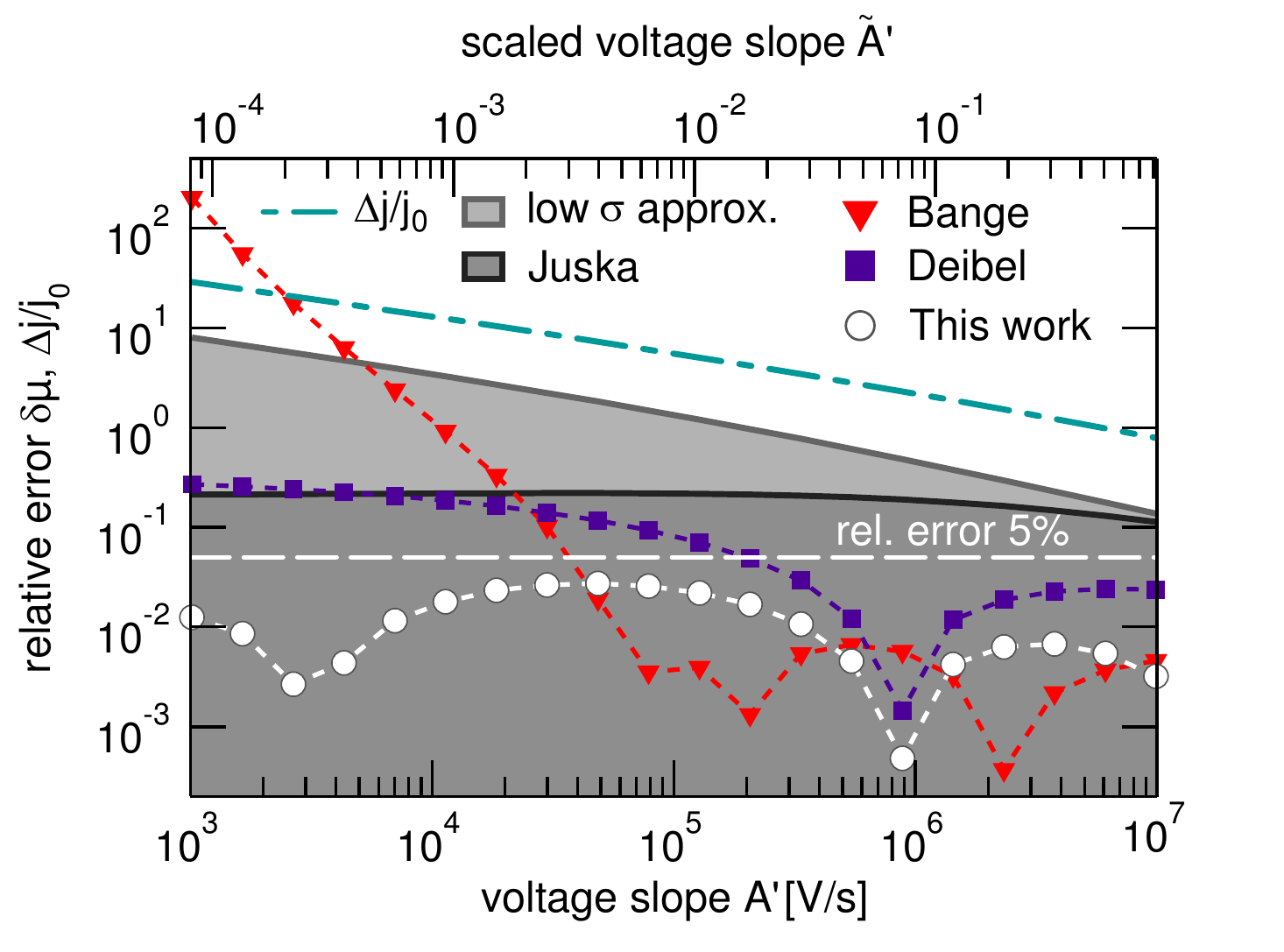}\hfill
  \caption{Comparison of the relative mobility error $\delta \mu$ for the different mobility equations from Ju{\v s}ka~\etal~\cite{juska2000a}, Deibel~\cite{deibel2009c}, Bange~\etal~\cite{bange2010} and \eqnref{eq:nlorr} together with the mobilities calculated for the low conductivity approximation from~\eqnref{eq:5a} versus the voltage slope $A'$. Parameters used to calculate the transients are: mobility $\mu=10^{-4}~\mathrm{cm^2/Vs}$, device thickness $d=150~\mathrm{nm}$, charge carrier density $n=6\cdot10^{22}~\mathrm{m^{-3}}$, dielectric constant $\epsilon=3.3$. The dashed line is added for orientation and represents an relative error of $5\%$. The dash-dotted line is $\Delta j/j_0$. Therewith we can relate the relative mobility error curves to the general case in \figref{fig:fig4}. The top axis holds the dimensionless voltage slope $\tilde{A}'$ for comparison. }
  \label{fig:fig5}
\end{figure}

  In case of the mobility equation~\eqnref{eq:mu_scale} parametric approximations for $\tilde{t}_{max}$ had been suggested in literature~\cite{juska2000,juska2000a,bange2010,deibel2009c}. The relative errors of which are shown in \figref{fig:fig5}. For low conductivities attended by small values of $\Delta \tilde{j}/\tilde{j}_0 \ll 1$, $\tilde{t}_{max}$ can be analytically derived as $\tilde{t}_{max} = \sqrt{1/3}$ ~\cite{juska2000}. However, for common experimental conditions this equation does not provide adequate accuracy and it remains impossible to analytically define $\tilde{t}_{max}$ in the general case. Thus, numerical estimated corrections have been predicted to account for the redistribution of the electric field. In Ref.~\cite{juska2000a,bange2010,deibel2009c} $\tilde{t}_{max}$  have the same type
  \begin{eqnarray}
    \tilde{t}_{max}=\sqrt{\frac{1}{3 ( 1 + \chi \frac{\Delta j}{j_0} )}}
    \label{eq:scalemu1}
  \end{eqnarray}
  with a correction factor $\chi$. Ju{\v s}ka~\etal~\cite{juska2000a} found $\chi=0.36$, Deibel~\cite{deibel2009c} suggested $\chi=0.21$ and Bange~\etal~\cite{bange2010} published $\chi=0.18$. Bange~\etal suggested an additional parametrization with a linear combination of two exponentials and four numerically derived adjusting parameters, which yields a very good fit for $\Delta j/j_0 < 7$~(Ref.~\cite{bange2010} Eq.~(4)). In~\figref{fig:fig4}~(b) and~\figref{fig:fig5} the curves denoted by ``Bange'' refer to the latter equation from Ref.~\cite{bange2010}. Anyhow, in~\figref{fig:fig4}~(b) none of the parametric approximations provides a good fit over the entire range of $\Delta j/j_0$. Accordingly, the best approximation we found giving reasonable results over the entire range of experimental parameters is
  \begin{eqnarray}
      \tilde{t}_{max}=0.5\left[\frac{1}{6.2 \left( 1 + 0.002 \frac{\Delta j}{j_0} \right)} + \frac{1}{\left( 1 + 0.12 \frac{\Delta j}{j_0} \right)}\right]
      \label{eq:scalemu}\mathrm{~.}
  \end{eqnarray}
  We clearly see in \figref{fig:fig4}~(b), that \eqnref{eq:scalemu} renders the calculated relation very well, instead all the other parametrization deviate sooner or later. Finally we substitute $\tilde{t}_{max}^2$ with \eqnref{eq:scalemu} in \eqnref{eq:mu_scale} and yield
  \begin{eqnarray}
    \mu=\frac{d^2}{2A' t_{max}^2}\left[\frac{1}{6.2 \left( 1 + 0.002 \frac{\Delta j}{j_0} \right)} + \frac{1}{\left( 1 + 0.12 \frac{\Delta j}{j_0} \right)}\right]^2
      \label{eq:mulorr}
  \end{eqnarray}
  for the mobility.

In \figref{fig:fig5} we show unscaled real parameters to compare the different mobility equations under experimental conditions. Therefore we have calculated photo-current transients within the CELIV framework with a defined mobility $\mu~=~10^{-4}~\mathrm{cm^2/Vs}$ and varied the applied extraction  voltage slope $10^{3}~\mathrm{V/s} \leq A' \leq 10^{7}~\mathrm{V/s}$. Plotted is the relative error $\delta \mu$ for the different mobility equations compared to the mobility $\mu$ we used as input parameter. To relate these values to the general case we added the variation of $\Delta j/j_0$ with varying voltage slope, as well as the top axes which represents the dimensionless voltage slope $\tilde{A}'$. Again our parametric mobility equation~(\eqnref{eq:mulorr}) is most suitable, as it yields an relative error which is smaller than $5\%$ over the entire range of experimental conditions.

\section{Results}

\subsection{Fitting CELIV experiments}
\label{sec:fitting}
\begin{figure}[t]
    \centering
    \includegraphics[width=.5\textwidth]{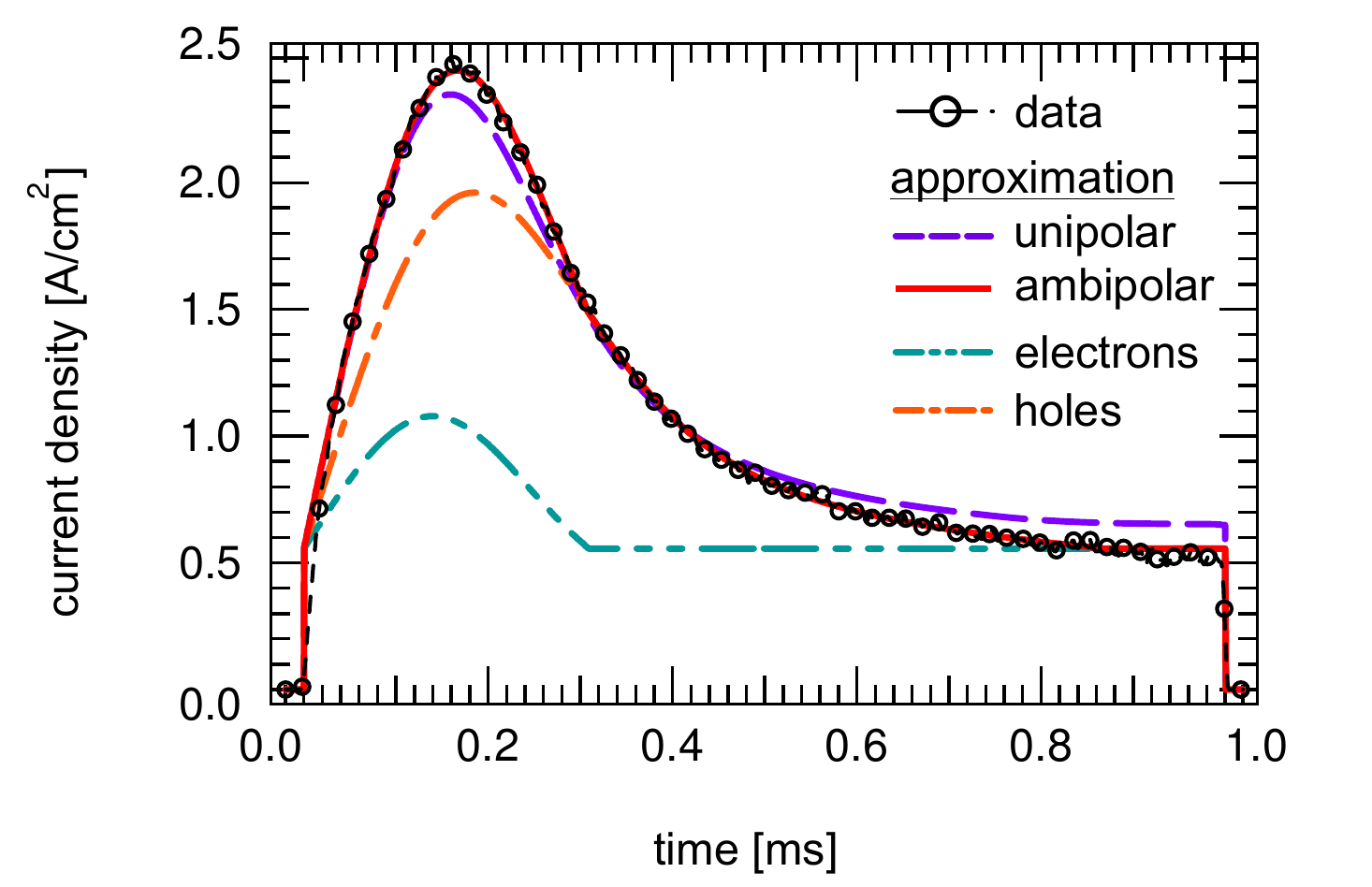}\hfill
    \caption{Results fitting photo-current transient from a \PCLV measurement at $T=180\mathrm{K}$ and a delay time $t_{delay}=20 \mathrm{\mu s}$. The black circles connected with a black thin dashed line correspond to the measurement. The thick dashed line represents the fit with a unipolar extraction current. For the solid line an ambipolar extraction is assumed for the fit. The two dash-dotted lines are the hole and the electron extracting currents, which superpose to the resulting photo-current (solid line).}
    \label{fig:fig3}
  \end{figure}

  In this section we briefly illustrate the computational determination of the charge carrier transport parameters from CELIV experiments by fitting them to our analytical framework. \figref{fig:fig3} shows a typical photo-current transient measured with \PCLV at $T=180~\mathrm{K}$ and a delay time $t_{delay}=20~\mathrm{\mu s}$ between laser excitation and extraction on a \pht:\pcbm bulk heterojunction solar cell. These measurements have already been presented by Deibel~\etal~\cite{deibel2008,deibel2009c} --- detailed experimental conditions and results can be found there.
  The parameters in \eqnref{eq:3} to be fitted to the experimental signal are the mobility $\mu$ and the charge carrier density $n$. The fitting results are summarized in Tab.~\ref{tab:tab1} together with the values determined from the measurement with~\eqnref{eq:mulorr} (mobility $\mu$) and~\eqnref{eq:nlorr} (charge carrier density $n$). We find that fitting with one extraction current (unipolar transport), taking one conducting type of charge carrier into account, fails, see \figref{fig:fig3}~(long dashed line). Here the fitted charge carrier density is too high and the time of the extraction current maximum $t_{max}$ is too short, thus the charge carrier mobility is overestimated. If we instead assume an ambipolar extraction to fit the photo-current, the resulting curve in \figref{fig:fig3}~(solid line) matches very well with the measured data. For the ambipolar extraction we used, as a first order approximation, a linear superposition of two extraction currents. The nature of the latter cannot be determined experimentally, but due to the photo-generation of excitons by laser excitation it is very likely to be hole and electron driven currents, respectively. As we see from the dash dotted and the dash double dotted lines in \figref{fig:fig3} the extraction current maxima are only slightly shifted with respect to each other, implying that the mobilities of holes $\mu_{h,fit}$ and electrons $\mu_{e,fit}$ are almost balanced, see Tab.~\ref{tab:tab1}. This is in a good agreement with time-of-flight measurements published in Ref.~\cite{baumann2008}. Correspondingly, we assign the slightly slower charge type as hole and the other one as electron transport to simplify the discussion. Furthermore the electron densities $n_{e,fit}$ are about a factor of two smaller compared to the hole densities $n_{h,fit}$ implying that the electron density is more strongly reduced within the delay time $t_{delay}$. Therefore  we propose the following two explanations without going too much into detail. Firstly, enhanced trapping of the electrons before the extraction seems possible. However, PCBM is found to be a trap-free acceptor~\cite{mihailetchi2003, mandoc2007, brabec2001review}, thus the trapping could not be energetically, but more a morphological effect, where the electrons are immobilized in isolated phases. For the small spherical fullerenes~\cite{hoppe2005} this is more likely than for the long polymer-chains. Secondly, in the \PCLV method an offset voltage is applied which compensates the built-in field to prohibit the extraction of charges before being swept out. However, the recombination of charge carriers changes the flat-band conditions continuously with time, thus a certain amount of charges is extracted during the delay time. Hence, electrons in conducting phases are more efficiently extracted due to their slightly higher mobility than holes. 
  \begin{table*}[t]
  \begin{tabular}{c c c c c c c}
  \toprule \hline
  & \multicolumn{3}{c}{ambipolar}    &unipolar    &\multicolumn{2}{c}{calculated}\\[0.05cm]
  \cline{2-4}\cline{6-7}
  & holes & electrons & mean/total & & integrated& this work\\
  & $x_{h,fit}$ & $x_{e,fit}$ & $x_{mean}$/$x_{total}$ & $x_{uni}$& $x_{int}$& $x_{calc}$\\[0.05cm]
  \hline
  mobility $\mu$ & \multirow{2}{*}{1.52} & \multirow{2}{*}{3.09} &
                  \multirow{2}{*}{2.30} &\multirow{2}{*}{3.69} &
                  \multirow{2}{*}{---} &\multirow{2}{*}{1.64}\\
  $\left(\times 10^{-6}\frac{cm^2}{Vs}\right)$ & & & & & &\\[0.05cm]

  density $n$ & \multirow{2}{*}{2.67} & \multirow{2}{*}{1.08} &
                  \multirow{2}{*}{3.75} &\multirow{2}{*}{7.68} &
                  \multirow{2}{*}{1.67} &\multirow{2}{*}{4.65}\\
  $\left(\times 10^{-16} cm^{-3}\right)$ & & & & & &\\[0.2cm]

  \hline \bottomrule 
  \end{tabular}
  \caption{Summary of the extracted parameters for the \PCLV measurement shown in~\figref{fig:fig3}. For the sake of clarity, the mobility $\mu$ and the charge carrier density $n$ is represented by the $x$ in the table header. The parameters are taken from the unipolar ($x_{uni}$) and ambipolar ($x_{h,fit}$,$x_{e,fit}$) fits or calculated ($x_{calc}$). The mobility is calculated from~\eqnref{eq:mulorr}, the charge carrier density from~\eqnref{eq:nlorr} and by integrating the extraction current density ($n_{int}$). The ambipolar weighted mean mobility is approximated by $\mu_{mean} = (n_{h,fit}\mu_{h,fit} + n_{e,fit}\mu_{e,fit})/(n_{h,fit}+n_{e,fit})$ and the ambipolar total charge carrier density is $n_{total}=n_{h,fit}+n_{e,fit}$.}
  \label{tab:tab1}
\end{table*}

  Finally, we combine in Tab.~\ref{tab:tab1} our fitting results ($\mu_{h,fit}$, $\mu_{e,fit}$, $\mu_{uni}$, $n_{h,fit}$, $n_{e,fit}$, $n_{uni}$) and the values we derived from the measurements with the help of \eqnref{eq:mulorr} ($\mu_{calc}$) and \eqnref{eq:nlorr} ($n_{calc}$). In addition, we define a weighted mean mobility $\mu_{mean} = (n_{h,fit}\mu_{h,fit} + n_{e,fit}\mu_{e,fit})/(n_{h,fit}+n_{e,fit})$ as well as the total charge carrier density $n_{total}=n_{h,fit}+n_{e,fit}$ from the ambipolar fit to compare it with the calculated values $\mu_{calc}$ and $n_{total}$. Furthermore, the charge carrier density is determined from the area below the photo-current transient ($n_{int}$), as it is commonly done~\cite{deibel2008,deibel2009c,mozer2005,dennler2006}. 
  
  From the \PCLV measurement we obtain a fitted hole mobility $\mu_{h,fit}$ comparable to the calculated $\mu_{calc}$. Instead, the weighted mean mobility $\mu_{mean}$ is slightly above $\mu_{calc}$ and the fitted electron mobility $\mu_{e,fit}$ is two times higher than $\mu_{calc}$. Thus, we assign the mobility determined from the photo-current peak maximum to the mobility of the holes, due to the higher hole density (see Tab.~\ref{tab:tab1}). In general the mobility $\mu_{calc}$ is related to the more conducting charge carrier type~\cite{mozer2005a}. From our results we conclude, that the the charge carrier density is the crucial parameter determining the charge carrier type, which is related to the mobility $\mu_{calc}$. Moreover, the extraction current~(\eqnref{eq:3}) depends quadratically on the charge carrier density and only linearly on the mobility.

  We note that the integrated charge carrier density $n_{int}$ accounts for the density of extracted charges, while the values from the transient fitting and from \eqnref{eq:nlorr} reflect the total density of mobile charges involved in the photo-current. As expected the density of extracted charges $n_{int}$ is smaller than the values derived from \eqnref{eq:nlorr} $n_{calc}$ and from the ambipolar fit $n_{total}$. About $65~\%$ of the photo-generated charge carriers could be extracted within the length of the applied voltage pulse $t_p=1~\mathrm{ms}$. Experimentally, this limitation is often observed, when the trade-off between the applied maximum voltage, the signal to noise ratio and the extraction time prevents the complete extraction of all free charge carriers. The tail slope of the extraction current and hence the amount of extracted charges gets very small. Thus even a two times longer extraction time is not capable to extract all charges. However, the total density of mobile charges derived from the ambipolar fitting $n_{total}$ and from the calculation via \eqnref{eq:nlorr} $n_{calc}$ are in a good agreement. Therefore and due to small root mean square deviation in \figref{fig:fig4}~(a), we recommend using \eqnref{eq:nlorr} to determine the charge carrier density $n$ from experiments.

  We want to point out that $n_{calc}$ should to be the upper limit of extracted charges. If this is not the case, the photo-current transients are distorted by processes which are not considered in the general CELIV framework, for instance recombination~\cite{bange2010} or trapping. In such a case the evaluation with any equation derived there from could be questioned. However, as shown in our measurements the extracted charge carrier density $n_{int}$ is smaller than the calculated $n_{calc}$.

\section{Conclusion}

  With the analytical solution for the extraction depth $l(t)$ we derived a complete framework for the CELIV technique. Therewith the extraction current response due to a linearly increasing voltage can be analytically calculated.

  We suggested two new parametric equations for the determination of the charge carrier mobility $\mu$ and the charge carrier density $n$ from the characteristics of CELIV transients, respectively. These equations are capable of handling the entire experimental range of parameters. The relative error of the charge carrier density equation does not exceed $2\%$ and accounts for the total density of mobile charges and not just for the extracted charge carrier density. Our mobility equation yields lower deviations from the analytical predictions than any previous suggested mobility equation.

  Finally we evaluated \PCLV measurements by fitting them within the derived analytical CELIV framework. We find that reasonably good fits can only be achieved, if an ambipolar extraction of holes and electrons is taken into account. The results show a balanced hole and electron mobility in P3HT:PCBM solar cells in accordance with previous experiments~\cite{baumann2008}. Furthermore we found that the type of charges with the higher charge carrier density, instead of the more mobile one, is mainly rendering the shape of the \PCLV transients and is therefore the one that is probed.

\section{Acknowledgements}
  The authors thank the Bundesministerium f\"ur Bildung und Forschung for financial support in the framework of the MOPS project (͑Contract No. 13N9867͒). C.D. gratefully acknowledges the support of the Bavarian Academy of Sciences and Humanities. V.D.'s work at the ZAE Bayern is financed by the Bavarian Ministry of Economic Affairs, Infrastructure, Transport and Technology.

\appendix
\section{The Airy functions}
  We briefly want to show how to normalize the Airy functions to yield the solution of x(t)~\eqnref{eq:8}. The Airy functions are defined for real values $x$ as follows
  \begin{eqnarray}
   Ai(x) &=& \frac{1}{\pi}\int_0^\infty \cos\left(\frac{t^3}{3} + xt\right)\mathrm{d}t\\
   Bi(x) &=& \frac{1}{\pi}\int_0^\infty \left[\exp\left(-\frac{t^3}{3} + xt\right) + \sin\left(\frac{t^3}{3} + xt\right)\right]\mathrm{d}t
   \label{eq:12}
 \end{eqnarray}
 and the values of $Ai(x)$ and $Bi(x)$ at $x = 0$ are
 \begin{eqnarray}
   Ai(0) &=& \frac{1}{3^{2/3}\Gamma\left(2/3\right)}\\
   Bi(0) &=& \frac{1}{3^{1/6}\Gamma\left(2/3\right)}
   \label{eq:13}\mathrm{~.}
 \end{eqnarray}
 The solution for \eqnref{eq:8} is the linear combination of $Ai(x)$ and $Bi(x)$
 \begin{eqnarray}
    x(t)&=& C_1 Ai'\left(\chi\right) + C_2 Bi'\left(\chi\right)\mathrm{~,}
  \label{eq:14a}\\
  \chi&=&\left(\frac{en\mu}{2\epsilon\epsilon_0d}\frac{\mu A'}{d} t^3\right)^{1/3}\label{eq:14b}\mathrm{~.}
 \end{eqnarray}
 We can derive the following boundary conditions for $x(0)$ and $x'(0)$ from $l(0) = 0$ at time $t = 0$
 \begin{eqnarray}
   x(0) &=& const.\\
   x'(0) &=& 0\mathrm{~.}
   \label{eq:15}
 \end{eqnarray}
  Therefrom we can determine $C_1$ and $C_2$ to
  \begin{eqnarray}
    C_1 &=& \frac{1}{2}3^{2/3}\Gamma\left(2/3\right)\label{eq:16a}\\
    C_2 &=& \frac{1}{2}3^{1/6}\Gamma\left(2/3\right)\label{eq:16b}\mathrm{~.}
 \end{eqnarray}

\section{Alternative definition of the extraction depth}
\label{hypergeo}
The Airy functions in \eqnref{eq:9} can be revealed by the confluent hyper-geometric function $_0F_1$. This yields a computational more robust definition of the extraction depth.
\begin{subequations}
\label{eq:17}
\begin{eqnarray}
    l(t)&=&\frac{\mu A'}{2d} t^2\frac{ {_0F_1\left[\frac{5}{3}, \frac{1}{9}\chi^3\right]}}{{_0F_1\left[\frac{2}{3},\frac{1}{9}\chi^3\right]}}\mathrm{,}
     \label{eq:17b}\\
     \chi&=& \left(\frac{en\mu}{2\epsilon\epsilon_0d}\frac{\mu A'}{d} t^3\right)^{1/3}\label{eq:17a}
\end{eqnarray}
\end{subequations}

\newpage 
\bibliographystyle{unsrt}
\bibliography{ref}

\end{document}